\documentstyle[prl,aps]{revtex}
\begin{document}
\title{Level Fluctuations and Many-Body Effects in Disorder-Free Quantum Dots}
\author{Min-Chul~Cha$^1$ and S.-R. Eric Yang$^{2,3}$}
\address{
1)Department of Physics, Hanyang University, Ansan 425-791,Korea \\
2)Department of Physics, Korea University, Seoul 136-701, Korea \\
3)Asia Pacific Center for Theoretical Physics,
Seoul, Korea}
\maketitle
\draft
\begin{abstract}
We have investigated whether many-body effects can induce
significant 
level fluctuations in a ${\it disorder}$-${\it free}$ quantum dot.  The closed 
energy shell structures and relaxation of the Hartree-Fock (HF) potentials
are found to play a significant role.
The level degeneracy consistent with the rotational symmetry of
the confining potential determines the structure of the energy shells.
A closed shell state of a dot can give rise to large fluctuations. 
When a strong magnetic field is present the shell structure is absent and
fluctuations are
significantly reduced.
\end{abstract}
\thispagestyle{empty}

Non-integrable mesoscopic systems exhibit Wigner-Dyson energy level
statistics that can be described by random matrix theory \cite{Me}.
Recently the ground state energy of quantum dots is found to 
display fluctuations considerably larger than those predicted
by random matrix theory \cite{Ashoori96,Chang96,Folk96,Sivan96}.
The interplay between disorder and electron-electron interactions 
is expected to be important 
in explaining the experimental data
\cite{Sivan96,Blanter97,Berkovits97,Tamura97}.  However, a deep
understanding of this problem is still lacking.
 
In a disorder-free dot single-particle effects will
not induce any fluctuations, and it   
is usually tacitly assumed that level fluctuations are significantly
large only in disordered systems.  
Whether adding many-body interactions in a disorder-free mesoscopic
system will produce observable fluctuations remains an open question.
This raises the fundamental question of what type of  
many-body effects can give rise to
significant level fluctuations in a quantum dot. 
Investigation of such fundamental problems may 
deepen our understanding the level
fluctuations of quantum dots. 

In this paper we propose such a many-body effect.
In a disorder-free quantum dot  
new energy shells can exist that are unrelated to the accidental 
degeneracy of the harmonic potential\cite{com}.
The energy shell structure is determined by the degeneracy reflecting the rotational 
symmetry of the confining potential, and will be present in 
any size of dot. 
Because of 2D rotational symmetry of a quantum dot  
the degeneracy of each shell is significantly
reduced compared to that of a real atom.  
The ordering of energy shells in the dot does not
follow a simple rule, and gives rise level fluctuations.  The physical origin of this effect 
is the relaxation of the HF potential,
which is omitted in the constant interaction model.
When an extra electron is added to the system the other electrons must readjust 
(relax) their positions self-consistently to minimize the total energy.
The new self-consistent potential may be significantly different
from the old one, and, therefore, energy ordering may change.  We find that this effect
can lead to ${\it significant}$ level fluctuations.
These mechanisms for fluctuations are consequences of
the interplay between the shape of the confining dot potential
and many body effects.

Our calculation is based on a Hartee-Fock approximation at zero temperature.
The HF single particle basis vectors \cite{com1} are chosen to be the eigenstates of
a quantum dot in a magnetic field described by the Hamiltonian
\begin{eqnarray} 
H_0 = {{\bf p}^2 \over 2m^*} + { 1 \over 2} m^* \omega^2 r^2
-{1 \over 2}  \omega_c {\bf L} \cdot {\bf \hat B}
-g \mu_B {\bf S} \cdot {\bf B},
\end{eqnarray}
where $m^*$, $B$, $\omega_c $ and $\omega$ are the effective electron mass,
magnetic field perpendicular to the 2D plane,
the cyclotron frequency, and the effective frequency 
$\omega^2=\omega_0^2+{1\over 4}\omega_c^2$.
A vector potential in a symmetric gauge is used here.
The operators $\bf L$ and $\bf S$ denote angular momentum
and spin of an electron.  
The eigenstate wave functions
of this Hamiltonian are labelled by orbital, angular momentum, and spin 
quantum numbers $n,l,s$:
\begin{eqnarray} 
\phi_{nls}({\bf r})={1 \over \sqrt{2\pi} a} e^{-il\theta}
R_{nl}({r^2 \over 2 a^2}) \chi_s
\end{eqnarray}
with
\begin{eqnarray} 
R_{nl}(x)=\sqrt{n! \over (n+|l|)!} e^{-x/2} x^{|l| \over2} L_n^{|l|}(x).
\end{eqnarray}
$\chi_s$ is spin wave function and the constant $a^2={\hbar / (2m^* \omega)}$.
The eigenenergies are 
\begin{eqnarray}
\epsilon^0_{n l s} = \hbar \omega (2n+|l|+1)-{1 \over 2} \hbar \omega_c l
-g \mu_B S_z B,
\end{eqnarray}
Note that in the presence of a magnetic field
$\epsilon^0_{n l s}\not=\epsilon^0_{n l -s}$.

The electron-electron interaction is given by
\begin{eqnarray}
V({\bf r}_i,{\bf r}_j)= 
{e^2 \over \kappa \sqrt{|{\bf r}_i - {\bf r}_j|^2 + d^2 }},
\end{eqnarray}
where $\kappa$ is the dielectric constant and
$d$ is the thickness of the planar quantum dot.
The HF Hamiltonian has a block diagonalized form. 
Each block may be labelled by $[l,s]$ and it has the following form
\begin{eqnarray}
D_{n n'}^{[l,s]}=<nls|H_0|n'ls>+<nls|V_H|n'ls>-<nls|V_X|n'ls>.
\end{eqnarray}
In order to find the matrix elements of the Hartree ($V_H$) and exchange ($V_X$) potentials
we need to
write HF eigenstates as a linear combination of the HF basis vectors
\begin{eqnarray}
|\alpha l s> = \sum_{m} c_{m}^{(\alpha l s)}|m l s>.
\end{eqnarray}
Invoking rotational invariance about the z-axis we may take 
the HF single-particle eigenstates,
$|\alpha l s>$, to be {\it eigenstates} of 
the z-component of angular momentum 
operator: $L_z|\alpha l s>=\hbar l|\alpha l s>$,
where $\alpha$ denotes HF orbital quantum numbers.
We find
\begin{eqnarray}
<nls|V_H|n'ls>=\sum_{\alpha,p,s'}<nls;\alpha p s'|V|n'ls;\alpha p s'>
f_{\alpha p s'}\nonumber\\
=\sum_{\alpha,p,s',m,m'}c^{* (\alpha,p,s')}_m c^{(\alpha,p,s')}_{m'}
<nls;m ps'|V|n'ls;m' ps'>f_{\alpha p s'}
\end{eqnarray}
and
\begin{eqnarray}
<nls|V_X|n'ls>=\sum_{\alpha,p}<nls;\alpha ps|V|\alpha ps;n'ls>
f_{\alpha ps}\nonumber\\
=\sum_{\alpha, p,m,m'}c^{* (\alpha, p,s)}_m c^{(\alpha, p,s)}_{m'}
<nls;m ps|V|m' ps;n'ls>f_{\alpha ps}.
\end{eqnarray}
where $f_{\alpha ps}$ are the Fermi functions.
Note also that even in the absence of a magnetic field the HF method allows 
partially spin-polarized solution.  In such a many-body state
$\alpha l+$ and $\alpha l-$ need not be degenerate.  
For a given number of electrons $N=N_++N_-$, the numbers 
of spin-up$(N_+)$  and -down $(N_-)$ electrons of the ground state
is determined by selecting $(N_+,N_-)$ which gives the lowest value of
ground state energy $E(N_+,N_-)$.
In such a state the Fermi levels for spin-up and -down
electrons match.

As in atomic physics it is possible to introduce energy shells in a quantum dot.
However, in a quantum dot the degeneracy of a shell is at most two,
consisting of states with two z-component of angular momentum quantum numbers $l$ and $-l$. 
A shell is thus denoted by $\alpha |l|s$ with 
$\alpha=1,2,...$ and $|l|=s,p,d,f,...$.
The two angular momentum states in a shell may not always be degenerate.
It is degenerate only when the total $L_z^{tot}=0$.  Our HF calculation shows that
states with $L_z^{tot} \not= 0$ can have lower energy than states with $L_z^{tot}=0$.
The total $L_z$ is zero when there are equal number of
{\it occupied} states with positive and negative signs of
angular momenta below the Fermi level for spin $s$.
The degeneracy is lifted when this number is unequal.
In such a case the exchange self-energies of the states $|\alpha,l,s>$ and $|\alpha,-l,s>$
are unequal because the exchange field couples states
with opposite signs of angular momentum differently:
$<\alpha,l,s|V_X|\alpha,l,s>\not=<\alpha,-l,s|V_X|\alpha,-l,s>$.

Like in a real atom the energy ordering of shell energies does not
follow a simple rule.  The physical origin of this effect 
is the relaxation of the HF potential.
If the single particle wave functions in the $(N+1)$- and $N$- particle 
systems are the same (no relaxation) energy ordering will be unchanged.  
However, as $N$ changes a new level may appear below the Fermi level:
For $N=8$ the spin-up (-down) ground configuration is $1s^11p^21d^2 (1s^11p^2)$. 
When an extra
electron is added a new level 2s  appears in the spin-up configuration 
$1s^11p^22s^11d^2$.

The interval between conductance peaks\cite{com2} is given by
\begin{eqnarray}
\Delta_N = \mu_{N+1}-\mu_N=E_{N+2} + E_{N} -2 E_{N+1}.
\end{eqnarray}
where the chemical potential $\mu_N=E_{N+1}-E_N$ and $E_N$ is the total  
ground state energy as a function of $N$.
$\Delta_N$ is plotted as a function of $N$ in Fig.1, where
$\hbar \omega_0 (\equiv \epsilon_0)
= 1 meV$, $\kappa=12.4$, $d=200 \AA$.
We remark the following features.  In the absence of a magnetic field 
$\Delta_N$ alternate between a trough and a peak as a function of $N$.
However, at
$N=13, 14$ and $N=16, 17$ this trend is broken and the values of 
$\Delta_N$ are almost unchanged.  The alternation of a trough and a peak 
reflects closed- and open-shell structures.
For $N=12$ the lastly occupied spin-down shell is $1d^1$, which is open (see Fig.2(a)). 
For $N=13$ the lastly occupied spin-up and -down shells are  
is $1f^2$ and $1d^2$, which are both
closed (see Fig.2(b)).  When the system is in a closed shell
state $\Delta_{N}$ is larger than the average value
while when it is in a partially filled shell $\Delta_{N}$ is smaller.
It is difficult to predict when $L_z^{tot}=0$ as a function of $N$ since the HF potential
relaxes in a subtle way as an extra electron is added.
The variance of $\Delta_N$ is $0.31\hbar \omega_0$.

In the presence of a magnetic field ($B= 1 T$)
$\Delta_N$ exhibits a  smooth dependence on  $N$, in contrast to the case of no magnetic field.
This different dependence on $N$ is due to the breaking of the rotational symmetry by
the magnetic field, i.e. the absence of closed shells.
The occasional interruption of the smooth dependence of $\Delta_N$
on $N$ is due to spin flipping:  For small $N$ Coulomb interaction dominates, 
and, consequently, to minimize the interaction energy, electrons tend to spin polarize.  
As $N$ increases the confinement energy dominates over the Coulomb energy 
and it is energetically favorable to pack electrons closely.
This state is achieved by flipping a spin and placing it at the center of the quantum dot \cite{Ya}. 
The numbers of spin-up and -down electrons are given in Table I for different
values of $N$.

We have also investigated the effect of random disorder. 
The random potential is described by
$H_{imp}({\bf r})= V_0 \sum_\alpha \delta({\bf r} - {\bf r}_\alpha)$
where ${\bf r}_\alpha$ are the positions of 400 randomly located impurities
in the area of $\pi (12 \sqrt{\hbar/2 m^* \omega_0 } )^2$,
and $|V_0| = 1.5$ meV with randomly chosen sign.
Figure 1 displays $\Delta_N$ and Table I shows the results of $(N_+,N_-)$ for the ground states.
Most of these results can be understood by invoking closed shell structures and relaxation of the HF
potentials.
However, due to the presence of disorder 
the degree of spin polarization in $(N_+,N_-)$
as a function of $N$ is changed in a subtle way.

The main result of our work
is that even in a disorder-free dot many-body effects can induce  significant level fluctuations.
Large level fluctuations do not seem to be a unique signature of a disordered dot.
The proposed mechanisms for level fluctuations may serve as a starting point
for the understanding level fluctuations
of disordered dots.

The work has been supported by the KOSEF under grant 981-0207-085-2, 
and by
Korea Research Foundation under grants
1998-015-D00128 and 1998-015-D00114.

\begin{figure}
\caption{Level spacing plotted as a function of electron number. Upper (lower) curves show results
in the absence (presence) of disorder.
The lower curves are shifted vertically for the sake of display.
The peaks in $\Delta_N$ reflect closed shell states of the dot.}
\end{figure}

\begin{figure}
\caption{Schematic energy diagrams for $N=12$ and $13$ (Energy is measured in units of 
$\epsilon_0=1 meV$). Unlike a real atom the maximum number of electrons that a level of a quantum dot can have is two.
Note that for $N=12$ spin-down (-up) shells $p$ 
and $d$ are non-degenerate (degenerate).  The explanation of this effect 
is given in the text.  As $N$ changes from 12 to 13 the ordering of $2s^1$ and $1d^2$ 
shells in the spin-up band is reserved
while in the spin-down band split $1p$ states become degenerate.}
\end{figure}

\narrowtext
\begin{table}
\caption{The number of spin-up($N_+$) and -down($N_-$) electrons  
for $\epsilon_0=1 meV$. 
In the absence of a magnetic
field the dot becomes spin-unpolarized as the electron number increases.
In the presence of a magnetic field the dot is almost completely spin-polarized.
For other values of $\epsilon_0$ the degree of spin polarization is different.}
\label{table2}
\begin{tabular}{ddddd}
   &{$B=0$ T}$\tablenotemark[1]$&{$B=1$ T}$\tablenotemark[1]$
   &{$B=0$ T}$\tablenotemark[2]$&{$B=1$ T}$\tablenotemark[2]$\\
$N$&$(N_+,N_-)$       &$(N_+,N_-)$
   &$(N_+,N_-)$       &$(N_+,N_-)$ \\
\tableline
 1&   (1,0)&  (1,0)  &  (1,0)  &  (1,0)\\
 2&   (1,1)&  (2,0)  &  (1,1)  &  (2,0)\\
 3&   (3,0)&  (3,0)  &  (3,0)  &  (3,0)\\
 4&   (3,1)&  (4,0)  &  (3,1)  &  (4,0)\\
 5&   (4,1)&  (5,0)  &  (3,2)  &  (5,0)\\
 6&   (5,1)&  (6,0)  &  (3,3)  &  (5,1)\\
 7&   (4,3)&  (7,0)  &  (6,1)  &  (6,1)\\
 8&   (5,3)&  (8,0)  &  (6,2)  &  (8,0)\\
 9&   (6,3)&  (9,0)  &  (6,3)  &  (9,0)\\
10&   (7,3)& (10,0)  &  (6,4)  & (10,0)\\
11&   (8,3)& (11,0)  &  (6,5)  & (11,0)\\
12&   (8,4)& (12,0)  &  (6,6)  & (12,0)\\
13&   (8,5)& (13,0)  &  (7,6)  & (13,0)\\
14&   (8,6)& (14,0)  &  (8,6)  & (14,0)\\
15&   (8,7)& (15,0)  &  (9,6)  & (15,0)\\
16&  (10,6)& (16,0)  & (10,6)  & (15,1)\\
17&  (10,7)& (17,0)  & (10,7)  & (16,1)\\
18&  (10,8)& (17,1)  & (10,8)  & (17,1)\\
19&  (12,7)& (18,1)  & (12,7)  & (18,1)\\
20&  (12,8)& (19,1)  & (12,8)  & (19,1)\\
21&  (13,8)& (20,1)  & (11,10) & (20,1)\\
22&  (12,10)& (20,2) & (12,10) & (20,2)\\
23&  (12,11)& (21,2) & (12,11) & (21,2)\\
24&  (12,12)& (22,2) & (12,12) & (22,2)\\
25&  (13,12)& (23,2) & (13,12) & (23,2)\\
26&  (14,12)& (23,3) & (14,12) & (24,2)\\
27&  (16,11)& (24,3) & (15,12) & (25,2)\\
28&  (16,12)& (25,3) & (14,14) & (25,3)\\
29&  (15,14)& (26,3) & (15,14) & (26,3)\\
\end{tabular}
\tablenotetext[1]{without disorder}
\tablenotetext[2]{with disorder}
\end{table}

\begin{references}
\bibitem{com}We only consider dots whose confinement potential 
is sufficiently weak such that
the magic degeneracy of harmonic potential( 2, 6, 12, 20,..) are lifted by electron-electron
interactions.
\bibitem{com1}
Our basis set consists of the 120 lowest energy states of a
non-interacting single-particle.
\bibitem{com2}
In HF approximation the fluctuation is given by
$\Delta_{N} = \epsilon_{N+1}^{(N+2)}-\epsilon_{N}^{(N+1)}$
where $\epsilon_{i}^{(j)}$ is the HF single particle
energy of eigenstate $j$ when $i$ electrons are
present.  
If the HF potential does not relax then  
$\epsilon_{i}^{(j)} \approx \epsilon_{i-1}^{(j)}+e^2/C$ 
and 
$\Delta_{N} \approx \epsilon_{N+1}^{(N+2)}-\epsilon_{N}^{(N+1)}+e^2/C$ ($e^2/C$ is the charging
energy).
\end{references}
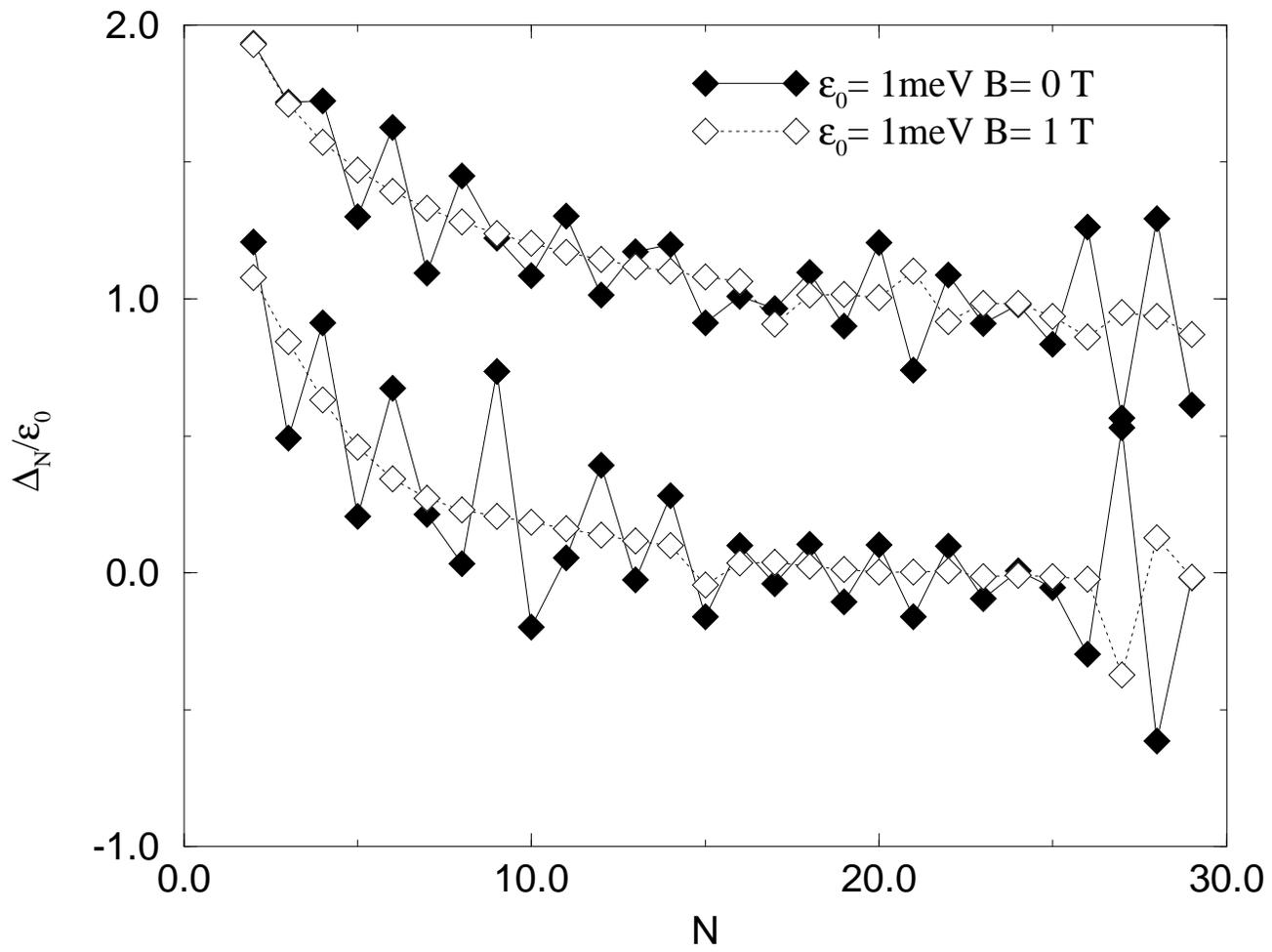
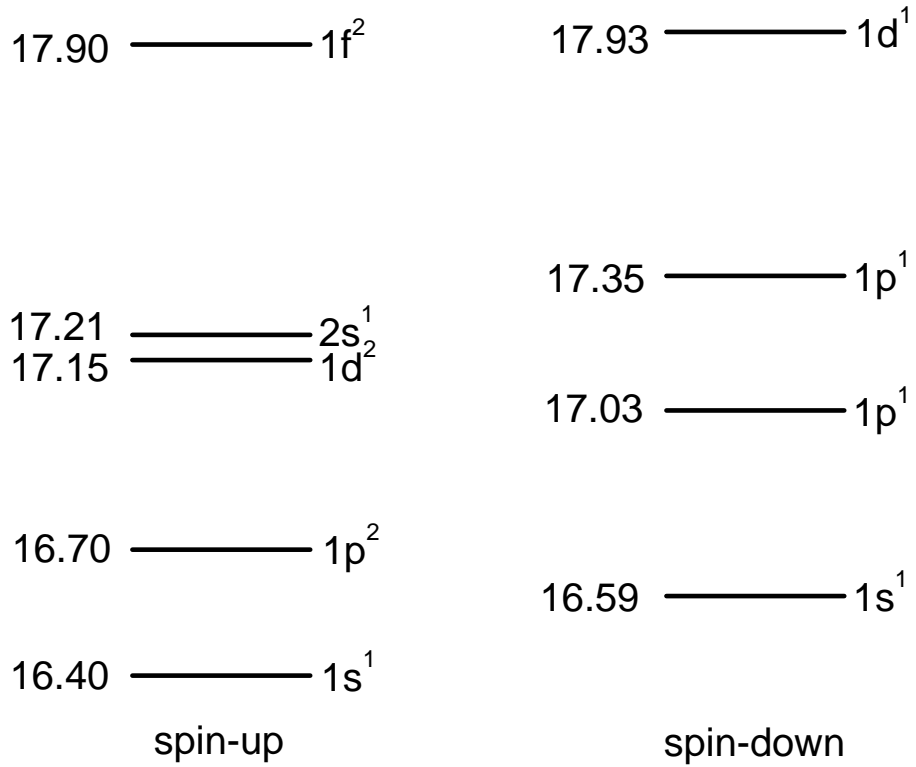
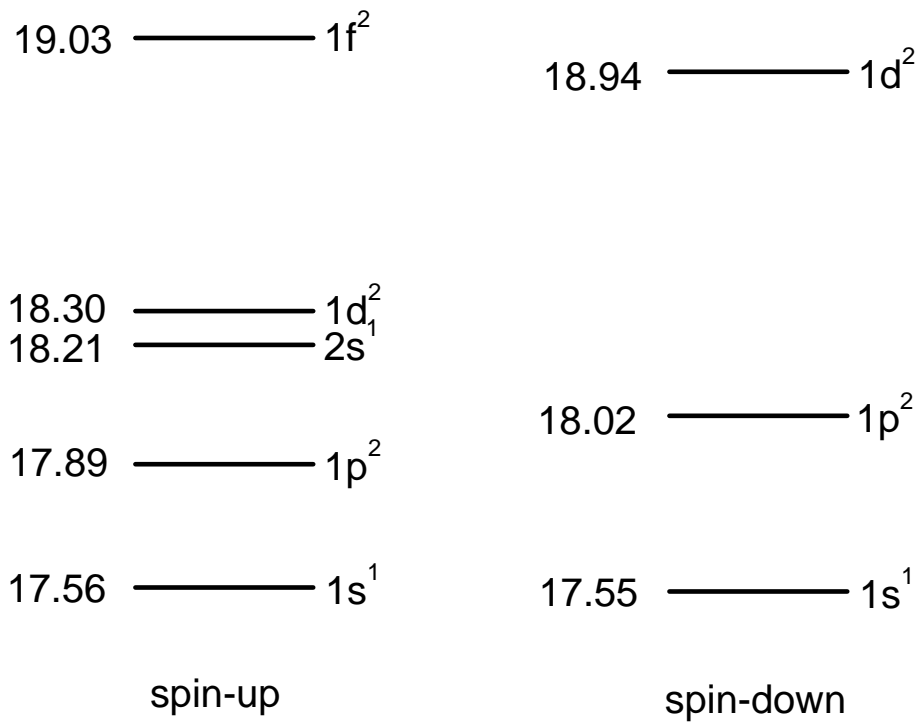
\end{document}